# A STUDY OF THE EFFECT OF VOLUME FRACTION ON STRESS TRANSFER WITHIN A UNIDIRECTIONAL FIBER REINFORCED COMPOSITE HAVING A BROKEN FIBER

Ajinkya V. Sirsat and Srikant Sekhar Padhee

Department of Mechanical Engineering, Indian Institute of Technology Ropar, Rupnagar, Punjab-140001

Keywords: FRC, RVE, STC, Randomness, Finite Element

## ABSTRACT

Stress concentration due to flaws in any material are very dangerous. It's understanding is thus very important before practical application of the material. As Fiber Reinforced Composite (FRC) is gaining wide range of application it is necessary to analyse it for stress concentration as one of the parameter. Literature survey reveals that the failure of FRC due to breakage of fiber is a cumulative process. If a fiber breaks stress concentration develops near the failure which leads to failure of other fibers in its vicinity. This process continues until the whole FRC gets failed. This phenomenon is quantified by a parameter called Stress Transfer Coefficient (STC). To analyse FRC generally it is practised to analyse Representative Volume Element (RVE) which is a representation of the complete FRC. Another important aspect which is considered while analysing FRC is the distribution of the fiber. Ideally the distribution of fibers should be uniform but no manufacturing process guarantees uniform distribution. Thus while analysing FRC it is a good practice to generate RVE with randomly distributed fibers.

In this paper an attempt is made to attain the relationship between volume fraction and STC. FRC with unidirectional fiber orientation is considered. RVE's with different volume fraction are analysed. A new method is implemented to generate RVE with randomly distributed fibers. RVE's are so generated that it contains a broken fiber at its geometrical centre and other fibers surrounding it. The RVE is loaded along the fiber direction.

## INTRODUCTION

Any structural design starts with material. To select proper material a prior information about it's properties is essential. Material characterization enables extraction of various material properties. The conventional material available are well characterized, but any new engineered materials needs proper characterization. With the advanced material the nature of characterization also needs to be updated. One such engineered material is composite. Composites are finding wide range of application in many broad areas like aerospace, automobiles, biomedical, etc. Fiber Reinforced Composites (FRC) needs to be characterized for the failure of fibers. From the literature it is clear that the failure of FRC due to breakage of fiber is a cumulative process [19]. If one fiber breaks in the composite there is stress concentration in the area of breakage. This stress is then transferred to the neighbouring fibers to maintain equilibrium [7]. This stress transfer may cause failure of other fibers leading to complete failure of FRC structure.

FRC are considered to have three structural levels: micro-scale, meso-scale and macro-scale [17]. The arrangement of fibers are considered in the micro-scale, the geometry of fiber/lamina is under meso-scale and the engineering structural response comes under the macro scale. Thus to characterize the FRC for the failure of fiber it is necessary to use the micro mechanical analysis approach. While performing the micro-mechanical analysis researchers have applied two kind of approaches. Many of them have considered the representative unit cell (RUC) approach [9] [12] [15] [16] [21] [22], the other have adopted representative volume element (RVE) approach [1] [10] [15] [17] [20]. RUC assumes that the fibers are arranged in periodic manner either in square array or in hexagonal array. The unit cell is chosen to be the smallest repeating periodic volume element [11]. RVE is considered to be the smallest volume of the microstructure which exhibits the same effective material properties that of the complete FRC [21]. Ideally the fibers should be distributed periodically, but in reality the fibers are never get arranged periodically rather they are randomly distributed. Thus the use of RUC will give poor characterization. Hence the RVE approach is used to do the micromechanical analysis of FRC with randomly distributed fibers.

The RVE should not be too large as it will be computationally expensive, it should also not to be too small that it cannot represent the FRC completely [8]. Thus, it is essential to select the proper size of the RVE. Trias et al. [15] have proposed that the size of the RVE should be fifty times the fiber radius. Grufman and Ellyin [21] proposed that the size of the RVE of 300μm x 300μm having approximately 400 fibers to completely analyse the microstructure. The statistically equivalent representative volume element (SERVE) was defined by Swaminathan et al. [13]. SERVE is defined as the smallest volume element of the microstructure which has the effective material properties and distribution function same as that of the entire microstructure. The SERVE can be chosen from any location within the entire microstructure.

To generate RVE with randomly distributed fiber many ideas and algorithms are proposed in the literature. Oh et al. [10] proposed an algorithm to generate RVE with random fiber arrangement up to volume fraction 60% and number of fibers equal to 120 to study the interfacial strain distribution under transverse loading. Wongsto and Li [18] generated RVE by disturbing the originally arranged hexagonal periodic array, in which he was able to get high volume fraction. Feder [4] has given random sequential adsorbtion (RSA) algorithm where the jamming limit of 54.7% was observed. Melro et al. [8] has proposed new algorithm which includes three step procedure: a) Hard core model, b) Stirring the fibers and c) fibers in the outskirts. Up to 500 fibers were able to be arranged without overlapping with this algorithm for volume fraction of 65%.Huang et al. [20] has studied the effects of fiber arrangements on mechanical behaviour while analysing the RVE generated by self-developed algorithm. Chateau et al. [3] used the algorithm which follows the model proposed by He et al. [6]. Another way to generate RVE was proposed by Zhang and Yan [21], they combined random disturbance algorithm and perfect elastic collision algorithm. In this method the fibers in the periodic arrangement was shifted to random arrangement by giving each fiber initial unit velocity in random direction. Bheemreddy et al. [2] generated a method which is a three step procedure: generation of square RVE followed by global crisscrossing and finally sub frame selection. Another approach of generating RVE is done by image processing technique. In which the realistic microstructure of FRC is characterized e.g. using scanning electron microscope (SEM) after that image processing is done on the image obtain from SEM, to locate the fibers in the microstructure. This data of fiber location is then used to generate statistically equivalent RVE imposing some algorithm. [17]

After developing the RVE micromechanical analysis was performed to carry out the required study. The current study is concerned with the characterization of the FRC for fiber failure which was inspired from the literature. Researchers are trying to characterize the FRC for the fiber failure from a long time. Hedgepeth [7] has proposed a shear lag model to analyse the effect of fiber breakage on other intact fiber. He has defined stress concentration factor (SCF) which quantifies the stress transfer which takes place in the FRC due to fiber breakage. Fukuda [5] has used local load shearing (LLS) model to analyse FRC in 2D. He concluded that the SCF does not follow any distribution though the fiber are distributed according to Weibull distribution. Swolf et al. [14] used the term stress transfer coefficient (STC) instead of SCF. With STC they also calculated the ineffective length and overload length. Zhao [22] used variational principles to calculate SCF and he compared the result with shear lag model which he found that they show good agreement. Yalin et al. [19] generated RVE with random fiber distribution using DIGIMAT-FE software and further analysed the RVE using Abaqus software. In the study the effect of distance to the broken fiber, effect of volume fraction and effect of fiber/matrix stiffness ratio was studied. In this paper a finite element model was generated using Abaqus. The model was altered to generate the fibers such that they make the compromise between the periodic arrangement and full random arrangement. The model was studied to observe the effect of distribution of the fibers on the STC. The effect of volume fraction on STC was also studied. The ineffective length was calculated for every volume fraction.

## MODELLING AND SIMULATION

**RVE:**

The model considered in the current study was inspired by Yalin et al. [19] model. The model has an RVE of 25μm X 25μm X 100μm. The model was further partitioned in 25 cells each of size 5μm X 5μm X 100μm. These cells are now filled with fibers, one fiber per cell. While filling the cells the fibers are generated randomly which follows normal distribution. Care was taken that the fibers do not cross the boundary of the cells. Thus a semi random distribution of fibers in the RVE was developed which is nothing but a compromise between the periodic square distribution and full random distribution. To generate the fibers PYTHON script was used. Figure 1 shows the typical RVE with the fiber distribution. The volume fraction of the RVE was changed by changing the radius of the fibers. Table 1 shows the fiber radius corresponding to the volume fractions and table 2 shows the material properties of matrix and fibers used in the analysis.

**Boundary Conditions and Meshing of the model:**

While modelling the RVE the fibers was taken of the length 100 μm except the central fiber which was taken to be 99μm. At z = 0 plane a displacement of 0.1μm was given in the direction outward normal to the plane. At z = 100 plane z-symmetric boundary condition was applied. The boundary conditions applied are shown in figure 2.

| Volume Fraction | Radius(μm) |
|---|---|
| 20 | 1.26 |
| 30 | 1.5 |
| 40 | 1.8 |
| 60 | 2.2 |

Table 1. Radius of fibers corresponding to the volume fractions.

| Material | Modulus | Poisson's Ratio |
|---|---|---|
| T800 carbon fiber | 394Mpa | 0.3 |
| Epoxy Resin | 3.4Mpa | 0.4 |

Table 2: Material properties of Matrix and fibers for the finite element model.

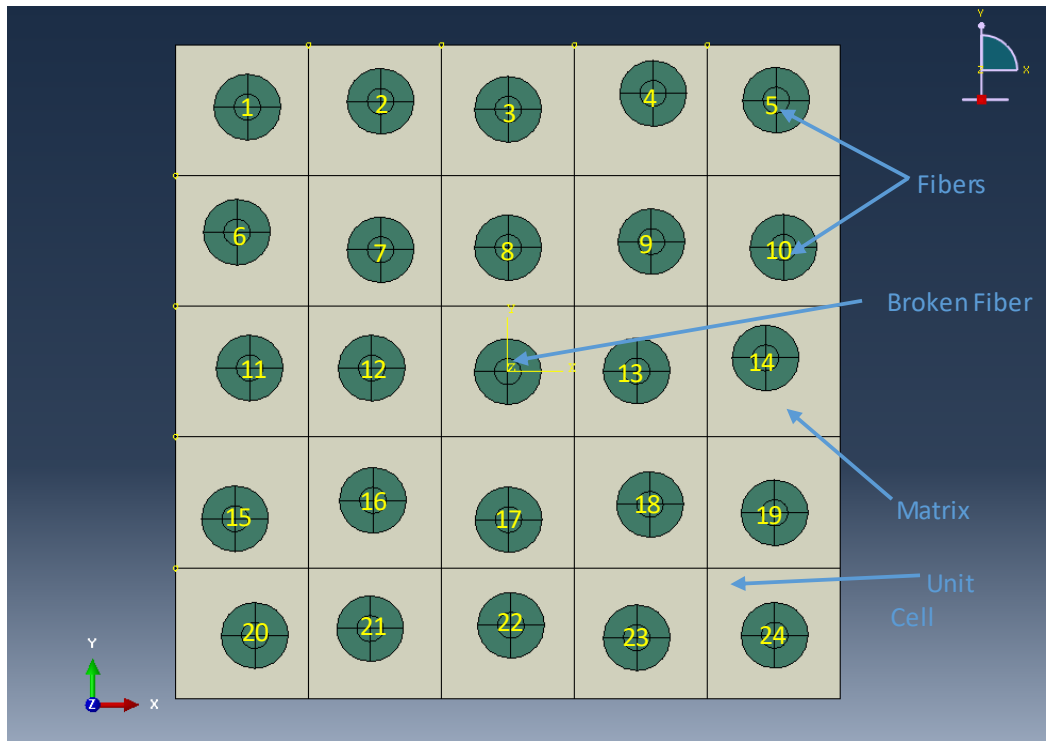

Figure 1. RVE showing the distribution of the fibers in the unit cell and the numbering of fibers.

For meshing the model, linear brick type 8- node (C3D8R) element was used. The number of nodes along x and y direction of the cells was taken to be 50. Along the length i.e. in the z direction the number of nodes are taken to be 33. The fibers was partitioned to have a number on nodes on the circumference equal to 64, also radially the fibers have nodes equal to 64. Thus in all the each fiber was meshed to have 1024 x 33 elements.

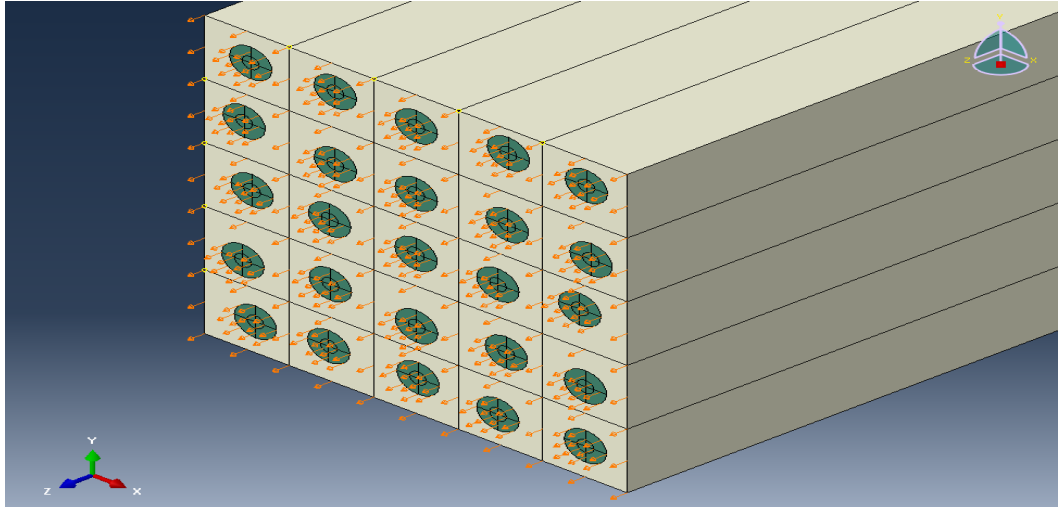

(a)

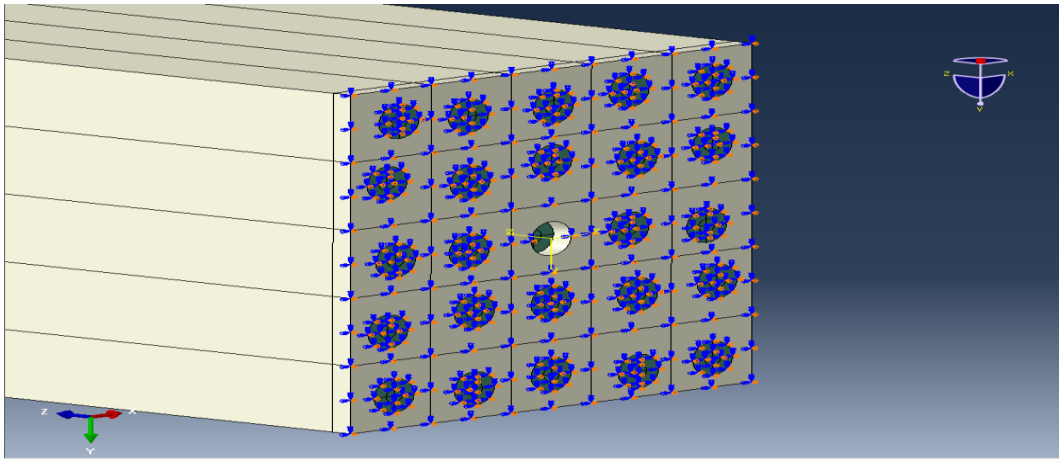

(b)

Figure 2. a) Displacement of 0.1µm given at z=0 and b) z-symmetry boundary condition is applied at z=100.

## RESULTS AND DISCUSSION

A python script was generated to automate the simulations in Abaqus. The script generates normally distributed random numbers which are used as the coordinates of the twenty-five fibers. For each volume fraction the script was repeated for 200 times to do the stochastic study of the distribution. STC which is defined as the ratio of the average stress in the vicinity of the breakage i.e. at z=100 to the average stress at far field i.e. at z = 0 was calculated for each iteration.

The STC was calculated for the nearest fibers and also for the second nearest fibers to the centrally broken fiber. Also ineffective length which is defined as twice the fiber length over which 90% of strain recovery occurs [14], is also observed for its variation with volume fraction and fiber distribution.

$$STC = \frac{\sigma *}{\sigma}$$

Where,

σ * = average stress in vicinity of fiber breakage

σ = average stress at far field

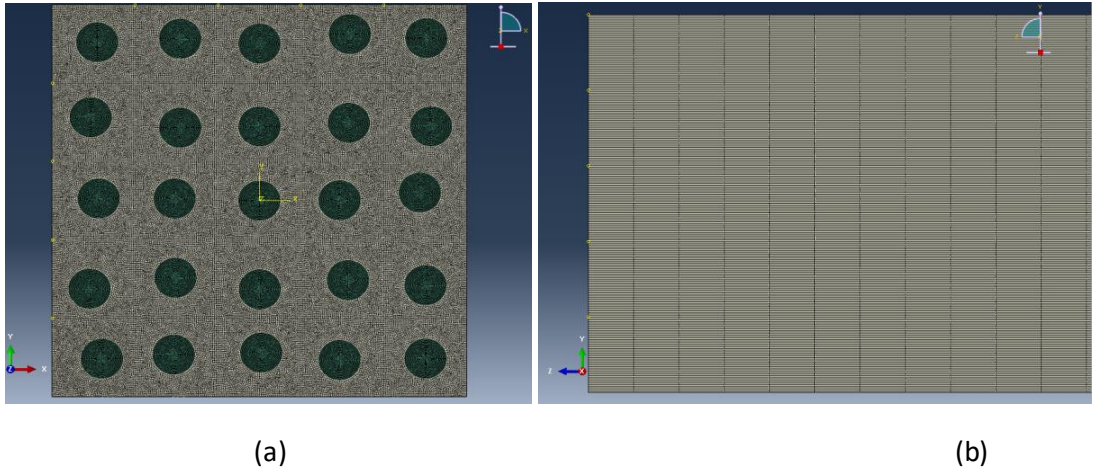

(a) (b)

Figure 3. a) Meshing of RVE in along z plane and b) meshing along the fiber direction

**Variation of STC due to fiber arrangement:**

Due to random distribution the fibers in some iterations fiber clustering is observed which results in high STC. Also in some cases the fibers are arranged away from the central fiber which result in low STC. The variation of the STC due to randomness in fiber arrangement is shown in figure 4. The most vulnerable fibers which may get break due to high stress are the neighbouring fibers numbered as 8, 12, 3 and 17. The next vulnerable fibers are the second nearest fibers which are named as 7, 9, 16 and 18. The numbering of fibers is shown in figure 1.

It is observed that the distribution of the STC due to the fiber arrangement does not follow strictly normal distribution, though the fiber distribution in each iteration follows the normal distribution. If closely observed for low volume fraction like 20% it can be seen that the distribution follows near to normal distribution because the radius of the fiber in these volume fraction is small as compared to the unit cell dimension, which gives proper space to the fiber to arrange more randomly. Similar kind of observation is observed in high volume fraction like 60 %. This is because the fiber radius in these case is not very small as compared to the dimension of the unit cell which provide less space for the fiber to get arranged more randomly. For volume fraction 30% and 40% it can be seen that the normal distribution is not at all followed by the STC.

**Variation of STC and Ineffective length due to volume fraction:**

As mentioned earlier the volume fraction of the RVE was changed by changing the fiber radius. From the result obtained after the simulations, it is observed that the volume fraction influences the STC. The values of STC increases with the increase in the volume fraction which can be seen from Table 3. As the volume fraction increases the fibers are getting closer to each other due to increase in the volume of the fiber, which means the volume of fraction is decreasing. Due to which the matrix is not able to release the stress within it and transfers large amount of stress to the surrounding fibers. The result of which being increase in the stresses in the surrounding fibers. The similar trend was observed by Swolfs et al. [14] and by Heuvel et al. [16].

The ineffective length does not show any significant change with the increase in the volume fraction.

| Volume Fraction (%) | STC |
|---|---|
| 20 | 1.0545 |
| 30 | 1.08 |
| 40 | 1.141 |
| 60 | 1.214 |

Table 3. Values of STC corresponding to the Volume fraction

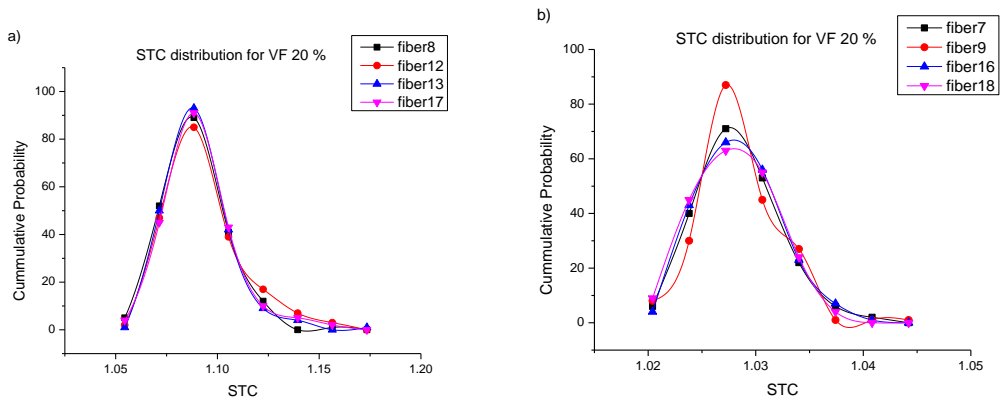

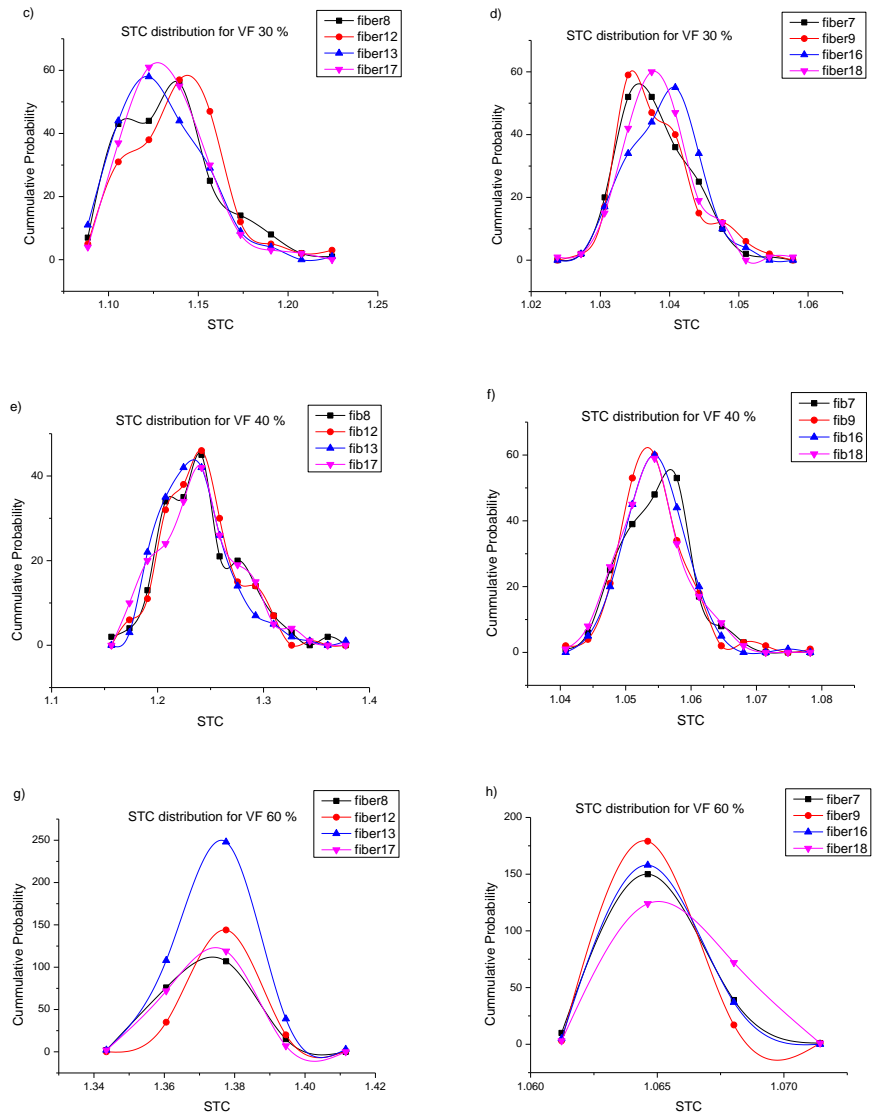

Figure 4. a,c,e,g) STC distribution in nearest fibers (fiber 8,12,13 and 17 for volume fraction 20%, 30%, 40% and 60% respectively and b,d,f,h) STC distribution for second nearest fibers(fiber 7,9,16 and 18) for volume fraction 20%, 30%, 40% and 60% respectively.

## CONCLUSION

A finite element study using Abaqus was performed on the RVE which has semi random fiber distribution. It is observed that the STC is influenced by the fiber arrangement. The volume fraction has a significant influence on the STC. It is observed that the STC increases with the increase in the volume fraction. For normal distribution of the fiber arrangement in RVE the STC is random however it does not follow normal distribution. Further it is interesting to observe that the ineffective length does not change with the fiber arrangement but it is influenced by volume fraction only.